\documentclass{appolb}
\usepackage{graphicx}
\usepackage{amsmath}

\begin{document}
\title{Application of Compressive Sensing Theory for the Reconstruction of Signals in Plastic Scintillators
  \thanks{Presented at Symposium on Applied Nuclear Physics and Innovative Technologies, 3-6 June 2013, Krak\'{o}w, POLAND}%
}
\author{L.~Raczy{\'n}ski$^{1}$, P.~Kowalski$^{1}$, T.~Bednarski$^{2}$, P.~Bia{\l}as$^{2}$, E.~Czerwi{\'n}ski$^{2}$, 
  {\L}.~Kap{\l}on$^{2}$, A.~Kochanowski$^{2}$, G.~Korcyl$^{2}$, J.~Kowal$^{2}$, T.~Kozik$^{2}$, W.~Krzemie{\'n}$^{2}$, 
  M.~Molenda$^{2}$, P.~Moskal$^{2}$, Sz.~Nied{\'z}wiecki$^{2}$, M.~Pa{\l}ka$^{2}$, M.~Pawlik$^{2}$, Z.~Rudy$^{2}$, 
P.~Salabura$^{2}$, N.~G.~Sharma$^{2}$, M.~Silarski$^{2}$, A.~S{\l}omski$^{2}$, J.~Smyrski$^{2}$, 
A.~Strzelecki$^{2}$, W.~Wi{\'s}licki$^{1}$, M.~Zieli{\'n}ski$^{2}$
\address{
$^{1}$~{\'S}wierk Computing Centre, National Centre for Nuclear Research, 05-400~Otwock-{\'S}wierk, POLAND \\
$^{2}$~Jagiellonian University, 30-059 Krak{\'o}w, POLAND\\
}
}

\maketitle
\begin{abstract}
Compressive Sensing theory says that it is possible to reconstruct a~measured signal if an enough sparse representation of this signal exists in comparison to the number of random measurements. This theory was applied to reconstruct signals from measurements of plastic scintillators. Sparse representation of obtained signals was found using SVD transform. 
\end{abstract}

\section{Introduction}

Medical imaging started in the end of 19th century when Roentgen discovered X-rays and has improved considerably since this time. Nowadays, examinations using imaging scanners are essential part of medicine. Thanks to these methods, pathologies can be observed directly rather than inferred from symptoms. For example, the growth of a brain tumor can be non-invasively monitored to determine an appropriate medical treatment. Medical imaging techniques can also be used when planning or even while performing surgery.

Positron Emission Tomography (PET) \cite{PET} is one of the most prominent and perspective techniques in the field of medical imaging. In this method the radiation is originated from inside of the patient’s body. Before an examination, the patient is injected with a drug containing a radionuclide, which localizes in a biologically active areas in the patient. In PET imaging a radioactive isotope emits positrons (anti-electrons). When a positron meets an electron, the collision produces a pair of gamma ray photons having the same energy at 511 keV but moving in opposite directions. The patient is surrounded by a large number of scintillation detectors and the two gamma quanta registered in coincidence define a line through the patient referred to as line of response (LOR). Using similar information from many positron annihilations, the image of distribution of the radionuclide is obtained.

Nowadays, in PET systems, a Time of Flight information is used to improve the image reconstruction process. In conventional PET, a positron annihilation is registered but its position along the LOR remains unknown. However, in Time of Flight PET (TOF-PET), the faster detectors are able to measure the difference in the arrival time of the two gamma rays, with the precision enabling to shorten significantly a range along the LOR where the annihilation occurred. The additional position information enables the reconstruction algorithm to improve the images quality and the procedure convergence.

In Strip TOF-PET scanner \cite{MOS1,MOS2,MOS3} gamma rays are detected in plastic scintillators. Signals propagate along the scintillators and are measured by two photomultipliers (PMTs) on its ends. Measurement of the arrival time at both PMTs allows to determine the place where gamma particle hit a~scintillator. Thus, the coincidence of hitting two scintillators is necessary to reconstruct a 3D position of positron annihilation.

In single strip of Strip TOF-PET system, time values are obtained by probing the signal in the amplitude domain. Probing must be used because there is no possibility to register a whole signals in time domain, since the duration of signals is given in nanoseconds. The information about the shape of the signals is highly correlated with the hit position of the annihilation gamma quantum, along the scintillator strip. Thus, with this information a better filtering of coincidence of the two signals and also a more accurate reconstruction of position are possible. 

But there remains a question: could we reconstruct the whole signal based on only a few random measurements? It is impossible to reconstruct the signal in the original time domain. However, according to the Compressive Sensing (CS) \cite{CS1,CS2} theory, the reconstruction is possible if an enough sparse representation of the signal exists in comparison to the number of random measurements.

The goal of this work was to investigate the possibilities of reconstruction of the original signal based on small number of measurements. In order to do this an appropriate representation of the data had to be found.

\section{Compressive Sensing theory}

Let us define the complete time signal and denote it with \textbf{\textit{y}}. The signal is \textit{K-sparse} if it is completely defined by only \textit{K} coefficients and \textit{K} is small in comparison to the signal dimension \textit{M}. The rest of the coefficients (\textit{M}-\textit{K}) are equal to zero or close to zero. The \textit{K-sparse} representation of the signal \textbf{\textit{y}}, will be denoted by \textbf{\textit{x}}. These coefficients may be calculated using Discrete Fourier Transform (DFT)~\cite{DFTandDWT}, Discrete Wavelet Transform (DWT)~\cite{DFTandDWT}, Singular Value Decomposition (SVD)~\cite{SVD} or any other transform. 

Since the signal \textbf{\textit{x}} is \textit{K-sparse} the reconstruction of its coefficients does not require the complete vector \textbf{\textit{y}}. The basic question in this scheme is how many random measurements of signal \textbf{\textit{y}} are needed to obtain a full information about the signal \textbf{\textit{x}}, and therefore full information about \textbf{\textit{y}}. In general, it can be shown that the \textit{N} random measurements of \textbf{\textit{y}} are required where \textit{N} is greater than \textit{2K} where \textit{K} is the sparsity of the \textbf{\textit{x}}. 

One can show, that it is possible to recover \textbf{\textit{x}} by solving an optimization problem of the form:
\begin{equation} 
  \widehat{\textbf{\textit{x}}} = arg(min\|\textbf{\textit{x}}\|_1), \:\:\:\:\:such\:that\:\textbf{\textit{y}} = \textbf{\textit{A}} \cdot \textbf{\textit{x}}
  \label{norm}
\end{equation}
where \textbf{\textit{A}} is the (\textit{M}x\textit{N}) transform matrix describing relation between \textbf{\textit{x}} and \textbf{\textit{y}}.
The \textbf{\textit{x}} and \textbf{\textit{y}} are \textit{M} and \textit{N} dimensional vectors, respectively. The 
$\|\textbf{\textit{x}}\|_1$ denotes the $L_1$ norm of vector {\textbf{\textit{x}}}. In general, the $L_p$ norm of \textit{M} 
dimensional vector {\textbf{\textit{z}}} is defined as follow:
\begin{equation}
  \|{\textbf{\textit{z}}}\|_p = \left\{ 
  \begin{array}{l l}
    \left( \sum\limits_{i=1}^{\textbf{M}}\vert z_i\vert^{p}\right)^{\frac{1}{p}} & p \in [1,\infty)\\
    max(\vert z_i\vert)& p = \infty 
  \end{array} \right.
\label{norm1}
\end{equation}

In case when $p < 1$, norm defined in Eq.~\ref{norm1} fails to satisfy the triangle ineqality, so it is 
actually a quasinorm.
The use of $L_1$ norm in minimization problem (see Eq. \ref{norm}) to promote or exploit sparsity has a long history 
\cite{L1MINa,L1MINb}. In general it can be shown that for $L_p$ norm, where \textit{p} is less or equal 1, sparse solution might be found.

\section{Signal representation}

According to Compressive Sensing theory, finding a proper representation of the data is very crucial for reconstruction of the signals.

\begin{figure}[htb]
\centerline{%
\includegraphics[width=9.0cm]{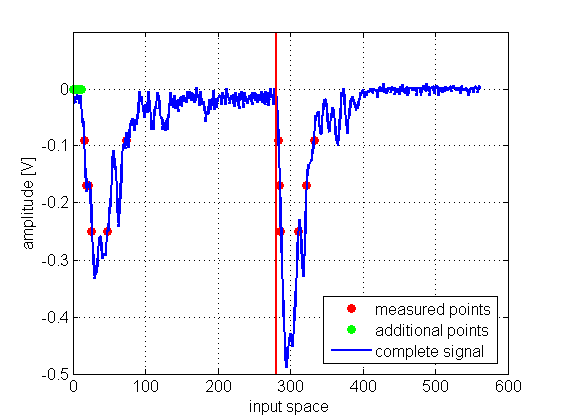}}
\caption{The example of a signal prepared for the reconstruction}
\label{SIGNAL}
\end{figure}

In defining the measured signal, denoted by \textbf{\textit{y}}, the information about a~time difference of arriving of the two signals is taken into account. For this purpose, the two independently registered signals from two detectors were concatenated. The example of signal \textbf{\textit{y}} is presented in Fig. \ref{SIGNAL}. 

The registration starts after one of the signals reach the amplitude of -0.05 V. In the example presented in Fig.\ref{SIGNAL}, the second signal arrived earlier. It can be assumed that before the registration starts, only a noise is present in the later signal (in an example a first one). Thus, in this case, the information about a time difference enable to put an additional points, marked with green dots. The amplitude values in these additional points are assumed to be equal to zero. The approach where whole the information from two signals is considered, facilitates the reconstruction.

Total number of points (\textit{N}) used for reconstruction of each \textbf{\textit{y}} signal, consists of twelve measured points (marked with red in Fig. \ref{SIGNAL}) plus number of points that express the difference between absolute time at both ends of scintillator (these additional points are marked with green dots). It should be noted that each signal \textbf{\textit{y}} consists of 560 samples.    

The complete data set used in this study consists of about 3500 signals. All signals in this data set are highly correlated. It seems reasonable to use a transform that builds basis vectors using data.

To extract the information present in the signals a decorrelation of the data set with the SVD transform was performed. The SVD transformation is commonly used to store the information more efficiently. The basis vectors of the SVD are the eigenvectors of the data set covariance matrix. 

\begin{figure}[htb]
\centerline{%
\includegraphics[width=9.0cm]{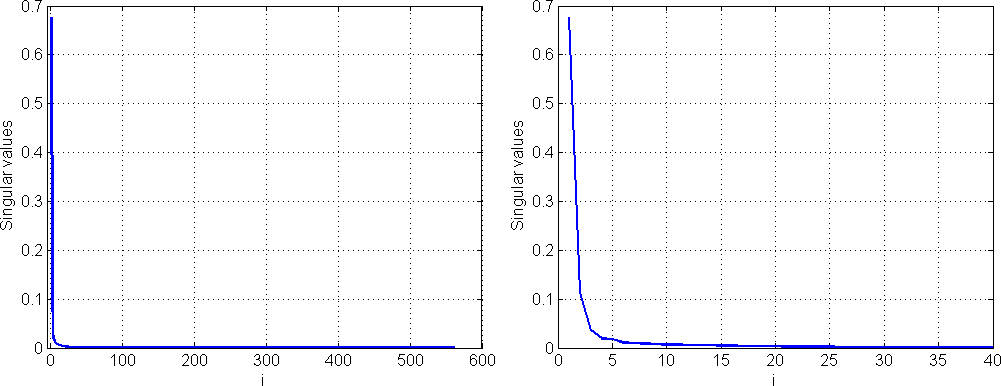}}
\caption{Singular values of dataset, with expanded view of first 40 values; $i$ denotes component number}
\label{SING}
\end{figure}

Fig. \ref{SING} shows the singular values of a 3500 x 560 data set. They were computed by first removing the column means of the dataset and then performing a full SVD. The singular values become very small for $i$ larger than 20. This means that the signal is sparse in the new SVD domain.

\begin{figure}[ht]
\centerline{%
\includegraphics[width=9.0cm]{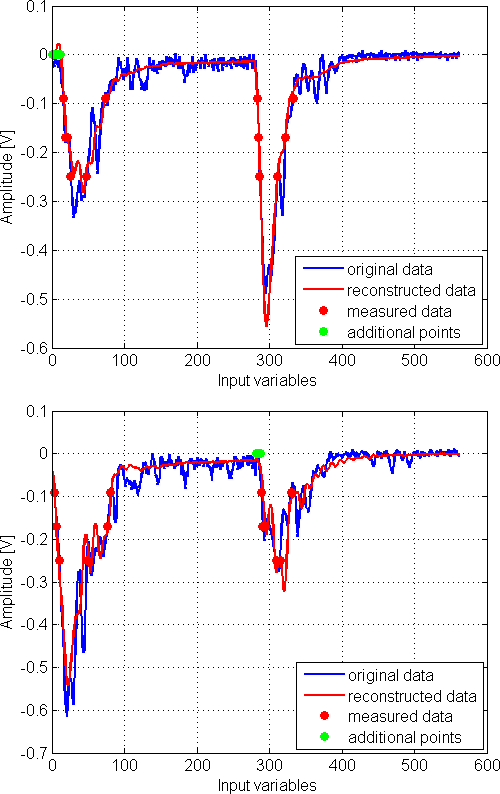}}
\caption{Example results of signal reconstruction}
\label{RESULTS}
\end{figure}

\section{Results}

The reconstruction of time domain signal (\textbf{\textit{y}}) is a two step process. In the first step, the sparse representation of signal in the SVD domain (\textbf{\textit{x}}) based on \textit{N} random measurements of \textbf{\textit{y}} is estimated. In the second step, the unknown signal \textbf{\textit{y}} is reconstructed based on estimated signal \textbf{\textit{x}} and the full SVD transformation matrix. The mean value of Root Mean Square Error (RMSE) between original and reconstructed \textbf{\textit{y}} signals from whole 3500 data set is c.a. 0.05 V. In a Fig. \ref{RESULTS} the two examples of original (blue curves) and reconstructed signals (red curves) are shown. 

In the upper image of Fig. \ref{RESULTS}, signal registered at second photomultiplier has higher amplitude what means that the place where gamma ray hit a~scintillator was nearer to the second PMT. The opposite situation is observed on the down image, where gamma ray hit a scintillator nearer to the first photomultiplier. In two presented examples the RMSE is similar and c.a. equal to mean RMSE from the whole dataset.

In the first example, only 21 points were used to reconstruct 560 samples of original signal. In the second example 560-sample signal was reconstructed using 18 samples.

\section{Summary}

In this paper a novel scheme for the reconstruction of signals in plastic scintillators was proposed. As it was shown, according to the Compressive Sensing theory, it is possible to reliably reconstruct full time signal using just several measurements. To achieve this result, a sparse representation of the data need to be found. In this work the transformation of the data based on Singular Value Decomposition (SVD) was examined. Reconstruction of signals from the photomultipliers using presented scheme is fast. It should be stressed that this is the case of using a SVD transform and for another transformation it could last much longer. The analysis of shapes of reconstructed signals may be used to reduce random coincidences and the background. The conception of the Compressive Sensing theory is very general and it may be used in solving other problems, where it is possible to find the sparse representation for the original representation of the data.\\

\textbf{Acknowledgments}\\
We acknowledge technical and administrative support by M.~Adamczyk, T.~Gucwa-Ry{\'s}, A.~Heczko, K.~{\L}ojek, M.~Kajetanowicz, G.~Konopka-Cupia{\l},
J.~Majewski, W.~Migda{\l}, and the financial support by the Polish National Center for Development and Research, the Foundation for Polish
Science through MPD programme, the EU and MSHE Grant No. POIG.02.03.00-00-013/09.


\begin{thebibliography}{abc}

  \bibitem{PET} J.~M.~Ollinger, J.~A.~Fessler, IEEE Signal Processing Magazine {\bf 14}, 43-55 (1997).
  \bibitem{MOS1} P. Moskal {\it et al.}, Bio-Algorithms and Med-Systems 14, Vol. {\bf 7},  73 (2011); e-print {\texttt{arXiv:1305.5187}}. 
  \bibitem{MOS2} P.~Moskal {\it et al.}, Nuclear~Med.~Rev. C {\bf 15}, 68-69 (2012); e-print {\texttt{arXiv:1305.5562}}.
  \bibitem{MOS3} P.~Moskal {\it et al.}, Nuclear~Med.~Rev. C {\bf 15}, 81-84  (2012); e-print {\texttt{arXiv:1305.5559}}.
  \bibitem{CS1} E.~Candes {\it et al.}, IEEE Trans. Inform. Theory {\bf 52(2)}, 489-509 (2006).
  \bibitem{CS2} E.~Candes {\it et al.}, Commun. Pure and Applied Mathematics {\bf 59(8)}, 1207-1223 (2006).
  \bibitem{DFTandDWT} Z.~Wang, Acoustics, Speech and Signal Processing, IEEE Transactions on {\bf{32.4}}, 803-816 (1984).
\bibitem{SVD} D.~P.~Berrar {\it et al.}, \textit{A Practical Approach to Microarray Data Analysis}, Kluwer Academic Publishers (2002).
\bibitem{L1MINa} A.~Beurling, {\it Ninth Scandinavian Mathematical Congress}, 345-366 (1938).
\bibitem{L1MINb} S.~Levy, P.~Fullagar, Geophysics, {\bf 46(9)} 1235-1243 (1981).

\end{thebibliography}
\end{document}